\documentclass[twocolumn,a4,prl,aps,floats,showpacs,amssymb,epsf]{revtex4}
\usepackage{graphicx}
\usepackage{bm}
\usepackage{amsmath}
\usepackage{afterpage}
\newcommand{\bfk}{\mathbf{k}}

\newcommand{\epsk}{\varepsilon_{\mathbf{k}}}
\newcommand{\epsF}{\varepsilon_{\mathrm{F}}}
\newcommand{\re}{\mathrm{Re}\,}
\newcommand{\im}{\mathrm{Im}\,}

\newcommand{\eV}{\,\mathrm{eV}}
\newcommand{\dl}[1]{\dot{#1}}
\DeclareMathAlphabet{\mathpzc}{OT1}{pzc}{m}{it}
\newenvironment{frcseries}{\fontfamily{frc}\selectfont}{}

\begin{document}

\title{Adsorbate-limited conductivity of graphene}

\author{John P. Robinson, Henning Schomerus,  L\'{a}szl\'{o} Oroszl\'{a}ny, and Vladimir I. Fal'ko}
\affiliation{Department of Physics, Lancaster University,
Lancaster, LA1 4YB, United Kingdom}

\date{\today}

\begin{abstract}
We present a theory of electronic transport in graphene in the
presence of randomly placed adsorbates. Our analysis predicts a
marked asymmetry of the conductivity about the Dirac point, as
well as a negative weak-localization magnetoresistivity. In the
region of strong scattering, quantum corrections drive the system
further towards insulating behavior. These results explain key
features of recent experiments, and are validated by numerical
transport computations.
\end{abstract}

\maketitle


Graphene (the two-dimensional allotrope of carbon~\cite{castronetoreview})
offers a chemically  stable platform to host various chemical
adsorbates~\cite{schedin,echtermeyer,bolotin,chen,lemme,chen2}. Their
presence strongly affects electronic transport, making graphene based devices
suitable for chemical sensoring. A key observation in experiments on
chemically functionalized samples is a marked asymmetry of the conductivity
as a function of a back gate voltage, which is used to steer the system
across the charge neutrality point (the Dirac point, which separates the
valence band from the conduction band). The conductivity becomes symmetric
only when the sample is annealed by a strong bias current which dislodges the
adsorbates.

\begin{figure}
\includegraphics[width=.95\columnwidth, bb=0 0 367 442]{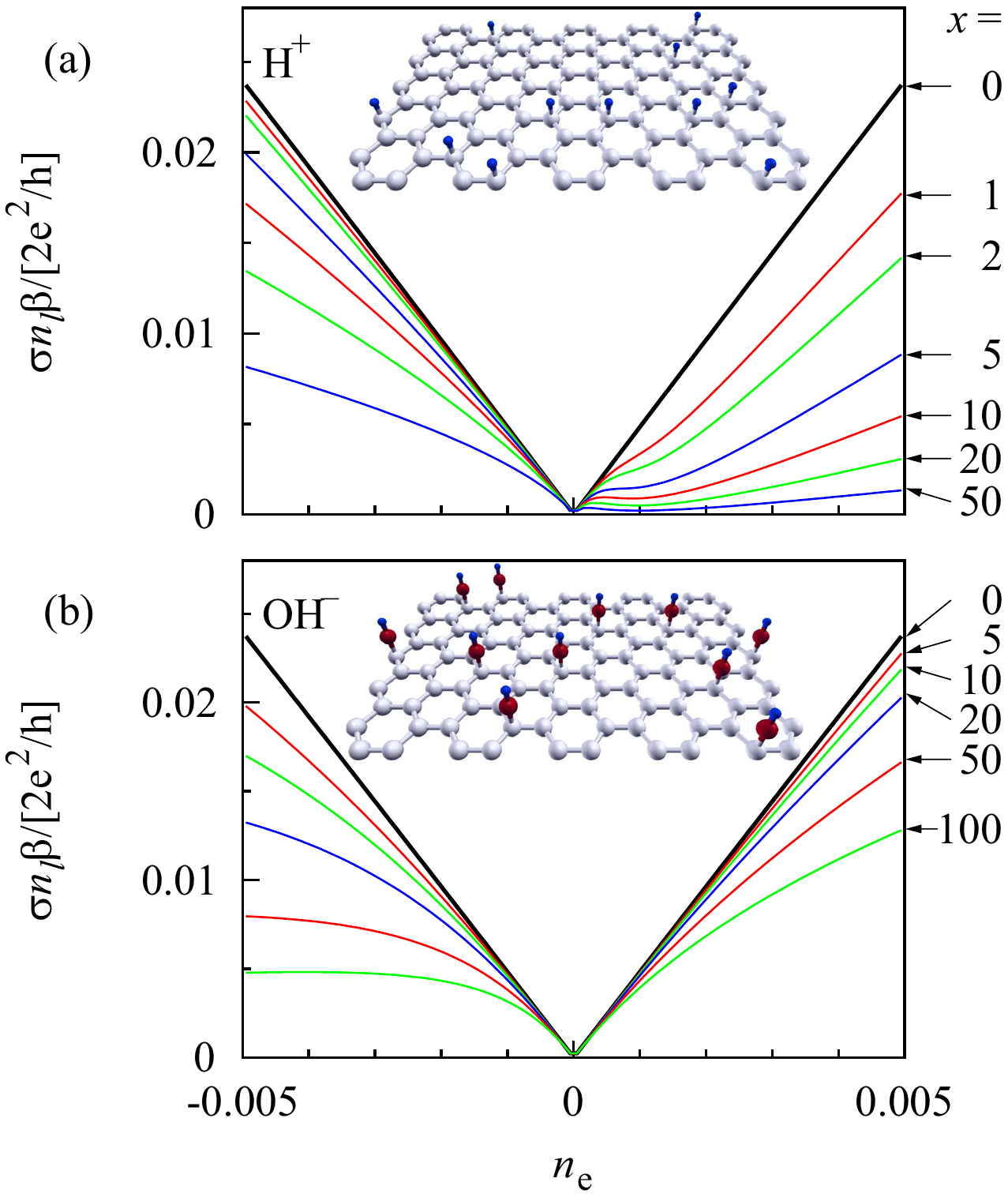}
\caption{(Color online) Conductivity of graphene in the presence
of adsorbates (concentration $n_i$) and Coulomb impurities
(concentration $n_l$) as a function of charge carrier
concentration  $n_e$ (carrier density $n_e/A_c$ with
$A_c=3\sqrt{3}a^2/4$ the area per carbon atom), for various ratios
$n_i/n_l=(\beta/2\pi) x$ where the dimensionless parameter
$\beta\simeq 1 $  characterizes the scattering strength of the
 Coulomb impurities \cite{nomura,hwang,remark}.
 (a) H$^{+}$, (b) OH$^{-}$, with
parameters from first-principle computations (the insets
illustrate sample segments). The results are based on Eq.\ (\ref{eq:sigmaall}).
} \label{fig1}
\end{figure}

In this paper we provide a theory, supported by numerical
simulations, that explains this experimental feature for
covalently bonded (chemisorbed) adsorbates
\cite{newref1,newref2,newref3}. Our model is based on the
tight-binding description of electrons in
graphene~\cite{dresselhausbook}. Chemisorbed molecules are
incorporated into this description as laterally attached
additional sites, where the onsite and coupling energies are
extracted from the band structure of a graphene sheet with
regularly placed adsorbates. Each type of adsorbate introduces a
characteristic local energy dependent scattering potential in the
graphene, and suppresses the conductivity on one side of the Dirac
point, with only a weak effect on the other. For example, adsorbed
H$^+$ yields almost insulating behavior in n-type graphene, while
in p-type structures the conductivity is close to that of clean
material. For OH$^-$ the role of the bands is reversed. When
combined with scattering from a random Coulomb
potential~\cite{nomura,hwang}, the resulting conductivity traces
(shown in Fig.\ \ref{fig1}) are consistent with the findings in
experiments~\cite{chen,lemme}. For a small adsorbate
concentration, these  conclusions can be drawn from kinetic
theory. For larger concentrations, we implement a recently
proposed renormalization group (RG) analysis~\cite{aleiner,mirlin}
to account for systematic quantum corrections to the conductivity
resulting from multi-adsorbate scattering. Our predictions are in
good quantitative agreement with the results of the numerical
transport computations.

In a graphene sheet, the carbon atoms are held together via sp$^2$-hybridized
covalent bonds, while the electronic transport takes place by hopping along
$\pi$ orbitals  which can participate in covalent bonding with adsorbates.
The electrons in the $\pi$-band of graphene with additional adsorbed atoms
can be described using a tight-binding Hamiltonian
\begin{subequations}
\begin{eqnarray}
&&\mathcal{H} =
-\gamma\sum_{\langle l,m \rangle} c^{\dag}_l
c_m + \sum_n{\cal H}_n, \label{eqn:H0} \\
&&{\cal H}_n = \varepsilon_id_n^\dagger d_n+\gamma_i(c_{\alpha_{n}}^\dagger
d_n+c_{\alpha_{n}}^{}d_n^\dagger). \label{eqn:Hn}
\end{eqnarray}
\label{eqn:Htot}
\end{subequations}
The first term of Eq.\ (\ref{eqn:H0}) corresponds to the Hamiltonian of clean
graphene, where $c_{l}$ are annihilation operators on sites of the honeycomb
lattice, and the nearest-neighbor coupling strength $\gamma=2\hbar v_D/3a$
(with bond length $a=1.42$~\AA) determines the Fermi velocity $v_D$ at the
Dirac point. Each adsorbed molecule contributes a term of the form
(\ref{eqn:Hn}), where $d_n$ is the annihilation operator on the adsorbate
site and $\alpha_n$ is the host position on the honeycomb lattice. The
adsorbate density is parameterized by $n_i/A_c$, where $n_i$ is the
adsorbate-to-carbon ratio, and $A_c=3\sqrt{3}a^2/4$ is the area per carbon
atom in graphene.

The model Hamiltonian (\ref{eqn:Htot}) can be justified by
first-principle calculations. Here we consider the adsorbates
H$^{+}$  and OH$^{-}$, chosen because of the presence of ambient
water in many experiments, and which also can be selectively
driven towards the graphene via electric fields
\cite{echtermeyer}. Using density-functional theory (DFT)
\cite{siesta}, we find that H$^+$ is described by
$\varepsilon_i=0.66 \,\gamma$, $\gamma_i=2.2\,\gamma$, while for
OH$^{-}$ $\varepsilon_i=-2.9\,\gamma$, $\gamma_i=2.3\,\gamma$.
These energies are comparable to the graphene hopping energy
$\gamma=2.6\eV$, which necessitates the nonperturbative approaches
employed in this paper. Our DFT calculations also confirm that the
adsorbates form covalent bonds to well-defined host sites. In the
fully relaxed configuration, both adsorbates are aligned in the
vertical direction of the graphene sheet (for illustration, see
insets of Fig.\ \ref{fig1}). Other energy scales (such as shifts
of the graphene onsite energies and next-to-nearest neighbor
couplings) are  small and can be safely neglected
\cite{jeloaica,katsnelson}.

The main building block of our analytical considerations is the
derivation of the scattering amplitude due to individual
adsorbates, which enters the collision term of kinetic theory and
also features as input into the renormalization group analysis
that captures quantum corrections due to multiple scattering.

In the first step we self-consistently eliminate the adsorbate sites from the
Hamiltonian (\ref{eqn:Htot}) via a decimation procedure. The electron
wavefunction can be written as $\left|\Psi\right\rangle=\sum_l \psi_l
\left|l\right\rangle + \sum_n\phi_n\left|n\right\rangle_{\rm
ad}\equiv|\psi\rangle\oplus |\phi\rangle$, where the amplitudes $\psi_l$ and
states $\left|l\right\rangle$ refer to the carbon sites, while $\phi_n$ and
$\left|n\right\rangle_{\rm ad}$ refer to the adsorbate sites. We now project
the Schr\"{o}dinger equation onto an adsorbate site, $\left\langle
n\right|_{\rm ad}(\varepsilon-\mathcal{H})\left|\Psi\right\rangle=0$ and find
that the amplitude
$\phi_n=\psi_{\alpha_n}{\gamma_i}/{(\varepsilon-\varepsilon_i)}$ on the
adsorbate is related to the amplitude of its host carbon site. The amplitudes
$\phi_n$ hence can be eliminated, which results in the reduced Hamiltonian
\begin{equation}
\widetilde{\mathcal{H}}= -\gamma\sum_{\langle l,m \rangle} c^{\dag}_l
c_m + \sum_n
Vc^{\dag}_{\alpha_n}c_{\alpha_n},\qquad V =
\frac{\gamma_i^2}{\varepsilon-\varepsilon_i},\label{barepotential}
\end{equation}
where the energy-dependent effective potential $V$ corresponds to the
self-energy which an adsorbate induces for electrons in graphene.

While the effective potential $V$ displays a distinct resonant
energy dependence, the analysis of the resulting conductivity
properties requires  combining this with the specific energetics
of the graphene sample, and in particular, with the existence of
the conical point at which the density of states drops to zero.
We therefore now turn to the analysis for the
Green's function $\mathcal{G}=(\varepsilon-\mathcal{H}+i0^+)^{-1}$
of the system.

For a single, well isolated adsorbate, the Green's function
can be obtained exactly over the entire energy range by utilizing
the $T$-matrix representation
\begin{equation}
\label{eqn:G} \mathcal{G}=\mathcal{G}_0 +
\mathcal{G}_0\mathcal{T}\mathcal{G}_0,\quad
\mathcal{T}\equiv(1-Vc^{\dag}_{\alpha}c_{\alpha}\mathcal{G}_0)^{-1}Vc^{\dag}_{\alpha}c_{\alpha},
\end{equation}
where $\mathcal{G}_0$ is the Green's function of the clean graphitic system and
$\mathcal{T}$ characterizes the scattering strength \cite{staticremark}. Expanding $\mathcal{T}$
in powers of $V\mathcal{G}_0$, and utilizing the translational and
crystalline symmetries of graphene (so that $\langle
n|\mathcal{G}_0|n\rangle\equiv g_0$ for all $n$), the resulting
series for $\mathcal{G}$ can be resummed yielding
\begin{equation}
\mathcal{T} = t_0(\varepsilon)c_{\alpha}^{\dag}c_{\alpha},\qquad
t_0(\varepsilon)=\frac{\gamma_i^2}{\varepsilon-\varepsilon_i-\gamma_i^2g_0(\varepsilon)}.
\label{eqn:T-matrix}
\end{equation}
In the latter expression, $t_0(\varepsilon)$  describes resonant scattering
of electrons in graphene from an adsorbate level renormalized by
hybridization with states in the $\pi$-band. This hybridization always shifts
the effective resonance level towards the Dirac point, which can be
understood as a consequence of level repulsion which pushes additional states
towards the region with the lowest density of states. The real part of
$\gamma_i^2g_0(\epsilon)$ [with $\re g_0(\varepsilon)=-\re
g_0(-\varepsilon)\approx (A_c/2\pi\hbar^2 v_D^2) \varepsilon
\ln(|\varepsilon|/\Delta)$, and $\Delta$  a high-energy cut-off] gives a
formal description of this systematic shift, while the imaginary part [with
$\im g_0=-\pi \nu_0(\varepsilon)$, and $\nu_0(\varepsilon)$ the density of
states per carbon atom] indicates a decrease of the resonance width near the
Dirac point.

\begin{figure}
\includegraphics[width=\columnwidth, bb=50 50 570 480]{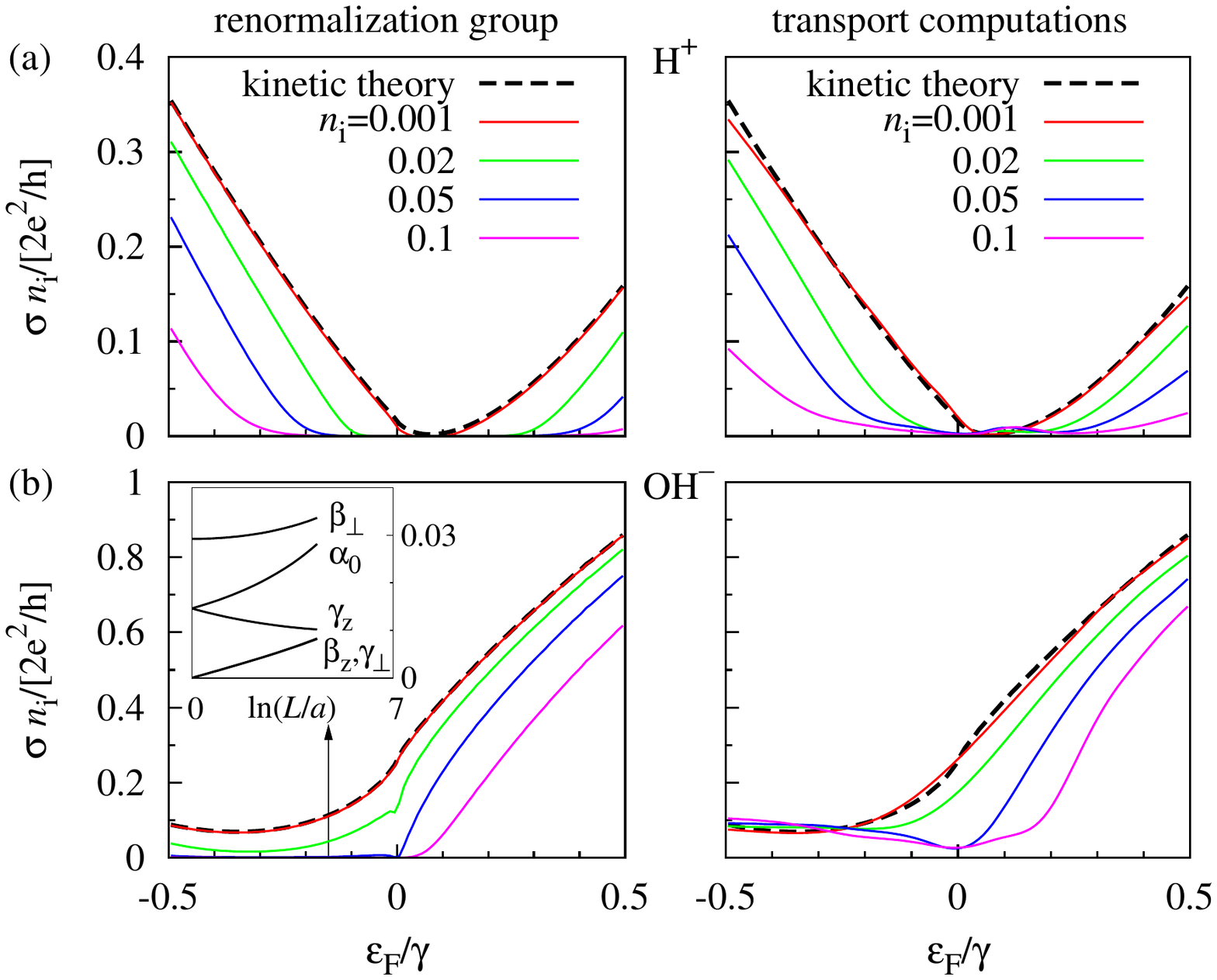}
\caption{\label{fig2} (Color online)  Conductivity of graphene in
the presence of adsorbates of variable concentration $n_i$. (a)
H$^+$, (b) OH$^-$. The dashed thick curve is the prediction
(\ref{eqn:sigma-drude}) of kinetic transport theory with
energy-dependent $T$-matrix (\ref{eqn:T-matrix}). The solid curves
in the left panels show the expected quantum corrections based on
the renormalization group analysis [Eqs.\ (\ref{eq:flow})]. In the
right panels, the solid curves show the results of numerical
transport computations.  The inset in the left panel of (b) shows
the RG flow (\ref{eq:flow}) of the dimensionless scattering
parameters for OH$^{-}$ at $\varepsilon=-0.15 \gamma$ and
$n_i=0.01$.}
\end{figure}

Ignoring (for the moment) the effects of phase-coherent multiple
scattering off the adsorbates, the conductivity of a disordered
sample with a finite adsorbate concentration $n_i$ can now be
obtained in kinetic transport theory. Due to $C_6$ rotational
symmetry of graphene, the conductivity is isotropic. Using the
scattering rate obtained from (\ref{eqn:T-matrix}),
\begin{equation}
\tau_{\bfk}^{-1} =\left(\frac{2\pi}{\hbar}\right)n_{i}|t_{0}(\epsk)|^{2}\nu_0(\varepsilon_{\bfk}),
\label{eqn:rate-1}
\end{equation}
where $\varepsilon_{\bf k}$ is the graphene dispersion relation,
we find the conductivity of graphene in the presence of the
adsorbates,
\begin{equation}
\sigma=\left(\frac{g_s
e^{2}}{h}\right)\frac{\hbar^{2}}{A_{c}n_{i}|t_{0}(\epsF)|^{2}}\frac{\langle
v_{\mathbf{k}}\rangle_{\epsF}}{\langle
v_{\mathbf{k}}^{-1}\rangle_{\epsF}}, \label{eqn:sigma-drude}
\end{equation}
where $\langle\ldots\rangle_{\epsF}$ denotes integration along the
Fermi line, $v_{\bf k}\equiv |\nabla_{\bf k}\varepsilon_{\bf
k}|/\hbar$, and $g_s=2$ accounts for spin degeneracy.

Motivated by recent experimental findings for selectively adsorbed
H$^+$ and OH$^-$ \cite{echtermeyer}, we show in Fig.\ \ref{fig2}
the corresponding dependence of the conductivity in Eq.~
(\ref{eqn:sigma-drude}) on the Fermi energy (thick dashed curve).
A key feature is the marked asymmetry of the conductivity about
the Dirac point caused by the energy dependence of the $T$ matrix
(\ref{eqn:T-matrix}). For H$^+$, over a range of energies in the
conduction band the conductivity is small, while it rises linearly
as one moves into the valence band, or far into the conduction
band. For OH$^{-}$ the  role of the bands is reversed.

In actual devices, adsorbate scattering is supplemented by
scattering off localized charges. These contribute to the
scattering rate a term $\tau_l^{-1}=n_l \beta
\gamma^2/(\hbar|\varepsilon|)$, where $n_l$ is the number of
localized charges per carbon atom, and $\beta\simeq 1$ is a
dimensionless numerical factor \cite{nomura,hwang}. For energies
in the linear part of the clean graphene dispersion relation, the
resulting conductivity can be written in the form
\begin{equation}
\sigma=\frac{2\pi\sqrt{3}}{n_l\beta}\frac{g_s e^{2}}{h}\left(x
|t_{0}(n_e)/\gamma|^{2} +n_e^{-1}\right)^{-1}, \label{eq:sigmaall}
\end{equation}
where $x=(2\pi/\beta) (n_i/n_l)$ characterizes the relative amount of the two
types of disorder, and $n_e=g_s\epsF^2/(2\sqrt{3}\pi\gamma^2)$ is the number
of charge carriers per carbon atom. Figure \ref{fig1} shows the charge
carrier dependence of the conductivity for various values of $x$. For $x=0$
the conductivity shows the symmetric linear charge-carrier dependence
characteristic for charged impurity scattering. For an increasing adsorbate
concentration, the conductivity develops the marked asymmetry discussed
above, while the Coulomb scattering only dominates very close to the Dirac
point, where it results in an additional dip.

Different types of disorder can also be discriminated via
mesoscopic corrections to the conductivity, which originate in
phase-coherent multiple scattering. In general, disorder in
graphene can be characterized in terms of five dimensionless
parameters $\Gamma=\{\alpha_0, \beta_{\perp}, \beta_z,
\gamma_{\perp}, \gamma_z\}$, which classify the breaking of the
symmetries  of the honeycomb lattice  \cite{falko}. Three of these
parameters describe intravalley scattering preserving the $C_{6v}$
symmetry ($\alpha_0$), the  $C_{3v}$ symmetry ($\gamma_z$), or no
point symmetry ($\gamma_\perp$), respectively. Analogously,
intervalley scattering is described by the parameters $\beta_z$
and $\beta_\perp$. In terms of these parameters, the conductivity
of graphene in kinetic theory takes the form
\begin{equation}
\sigma = \frac{g_s e^2}{\hbar\pi^2}\left(\frac{\alpha_0}{2} +
\beta_{\perp} + \gamma_\perp+ \frac{3}{2}\beta_z +
\frac{3}{2}\gamma_z\right)^{-1}.\label{eqn:sigma_RG}
\end{equation}
The scattering potential for adsorbates is such that
\begin{equation}
\alpha_{0}=\gamma_{z}=\beta_{\perp}/2=\frac{A_{c}n_{i}|t_{0}(\epsF)|^{2}}{2\pi\langle
v_{\mathbf{k}}\rangle_{\epsF}/\langle
v_{\mathbf{k}}^{-1}\rangle_{\epsF}},\quad
\beta_z=\gamma_\perp=0.\label{eqn:alpha0}
\end{equation}
This puts chemically functionalized graphene into the so-called
orthogonal symmetry class, for which one expects a negative
weak-localization magnetoresistance \cite{falko}, as well as
strong Anderson localization when the adsorbate concentration
increases \cite{aleiner}.

These corrections to kinetic theory can be studied systematically via a
renormalization group analysis \cite{shankar}, which provides effective
scattering parameters $\widetilde \Gamma$ that replace the bare values in
Eq.\ (\ref{eqn:sigma_RG}). For disorder representative of the symmetries in
graphene, the flow equations were derived in Ref.\ \cite{aleiner}. An
equivalent formulation in terms of the parameters $\Gamma$ given above can be
found in Ref.\ \cite{mirlin}, and takes the form
\begin{eqnarray} \dl{\alpha_0} &=&
2\alpha_0(\alpha_0 + \beta_{\perp} + \gamma_{\perp}
+ \beta_z + \gamma_z) + \beta_{\perp}\beta_z + 2\gamma_{\perp}\gamma_z , \nonumber \\
\dl{\beta_{\perp}} &=& 4(\alpha_0\beta_z + \beta_{\perp}\gamma_{\perp} + \beta_z\gamma_z) , \nonumber \\
\dl{\beta_{z}} &=& 2(\alpha_0\beta_{\perp} - \beta_z\alpha_0 + \beta_{\perp}\gamma_z + \beta_z\gamma_z) , \nonumber \\
\dl{\gamma_{\perp}} &=& 4\alpha_0\gamma_z + \beta_{\perp}^2 + \beta_z^2 , \nonumber \\
\dl{\gamma_z} &=&  2\gamma_z(-\alpha_0 -\beta_{\perp} + \beta_z +
\gamma_{\perp} -
\gamma_z) + 2\alpha_0\gamma_{\perp} + \beta_{\perp}\beta_z , \nonumber \\
\dl{\varepsilon} &=&  \varepsilon(1+\alpha_0 + \beta_{\perp} +
\gamma_{\perp} + \beta_z + \gamma_z), \label{eq:flow}
\end{eqnarray}
where $\dl X\equiv dX /d \ln (L/a)$ and $L$ is a running length. This RG flow
is integrated using the bare parameters $\Gamma$ and $\varepsilon=\epsF$ as
initial conditions and terminated when $\varepsilon$ reaches a high-energy
cutoff $\varepsilon_c$. We here implement this procedure for the
energy-dependent initial conditions (\ref{eqn:alpha0}). A typical solution of
the flow equations as a function of $\ln (L/a)$ is shown in the inset in
Fig.\ \ref{fig2}(b) (left panel). The dominance of $\beta_\perp$ over
$\gamma_z$ and $\gamma_\perp$ demonstrates that chemically functionalized
graphene stays in the orthogonal symmetry class when the adsorbate
concentration is increased. As shown  by the curves in the left panels of
Fig.\ \ref{fig2}, over the whole energy range and for both types of
adsorbates (H$^+$ and OH$^-$) the renormalization leads to a suppression of
the conductivity and drives the system towards insulating behavior.

In order to verify these expectations we carried out  numerical transport
computations, which are directly based on the model Hamiltonian
(\ref{eqn:Htot}). The conductivity $\sigma$ is obtained by finite-size
scaling of the Landauer conductance $G=\sigma W/L$ of graphene ribbons with
respect to width (in the range $50<W/\sqrt{3}a<100$) and length (in an
adaptively chosen range that avoids the onset of Anderson localization
\cite{schomerus}). For each geometry the conductance is computed in an
efficient recursive Green's functions algorithm \cite{john}, and averaged
over $10^4$ disorder realizations. The resulting dependence of the
conductivity on the Fermi energy (right panels of Fig.\ \ref{fig2}) is in
good agreement with the expectations based on the RG analysis (left panels of
Fig.\ \ref{fig2}). In particular, the numerical results clearly confirm the
suppression of the conductivity due to quantum corrections.

In summary, we have presented a theory of electronic transport for graphene
in the presence of chemisorbed molecules. We find that each type of adsorbate
results in a characteristic Fermi-energy dependence of the conductivity,
which is asymmetric about the charge neutrality point and distinguishes p-
and n-type transport. In a range of energies, the conductivity is strongly
suppressed, which could be used to increase the on-off ratio in
graphene-based field-effect transistors. These effects are further enhanced
by quantum corrections driving the system towards the localized state at
higher adsorbate concentrations. In experiments, the adsorbate concentration
can be increased by variable deposition times, or by driving adsorbates to
the sample using top gates \cite{echtermeyer,chen,lemme}, while the
localization effects can be probed via magnetoresistivity experiments on
graphene flakes at low temperatures \cite{wlexperiment}.

\acknowledgements

We wish to thank Tim Echtermeyer for providing us with
experimental data supporting the conclusions of this work, and
gratefully acknowledge discussions with I. Aleiner, C. Lambert, A.
Mirlin, and E. McCann. This work was supported by the European
Commission, Marie Curie Excellence Grant MEXT-CT-2005-023778, and
by the EPSRC.


\begin{thebibliography}{99}
\bibitem{newref1} Conduction asymmetries can also
arise for ionic bonding \cite{chen2,newref2} and substitutional
doping  \cite{newref3}.
\end{thebibliography}
\end{document}